\documentclass[fleqn,usenatbib]{mnras}

\usepackage{newtxtext,newtxmath}

\usepackage[T1]{fontenc}
\DeclareRobustCommand{\VAN}[3]{#2}
\let\VANthebibliography\thebibliography
\def\thebibliography{\DeclareRobustCommand{\VAN}[3]{##3}\VANthebibliography}

\usepackage[utf8]{inputenc}
\usepackage{graphicx}	
\usepackage{amsmath}	
\usepackage{amssymb}	
\usepackage{url}

\newcommand{\mr}[1]{\mathrm{#1}}

\title[Hot ultramassive rapidly rotating DBA White Dwarf]{Discovery of a hot ultramassive rapidly rotating DBA White Dwarf}

\author[M. Pshirkov et al.]{
M. S. Pshirkov$^{1,2}$\thanks{E-mail:pshirkov@sai.msu.ru},
A. V. Dodin$^{1}$,
A. A. Belinski$^1$,
S. G. Zheltoukhov$^{3,1}$,
A. A. Fedoteva$^{3,1}$,
\newauthor
O. V. Voziakova$^1$,
S. A. Potanin$^{1,3}$,
S. I. Blinnikov$^{4,1}$
and K. A. Postnov$^{1,4}$
\\
$^{1}$ Sternberg Astronomical Institute, M.V. Lomonosov Moscow State University, 13, Universitetskij pr., 119234, Moscow, Russia\\
$^{2}$
Institute for Nuclear Research of the Russian Academy of Sciences, 117312, Moscow, Russia\\
$^{3}$ Faculty of Physics, M.V. Lomonosov Moscow State University, Leninskie Gory, 1, 119991, Moscow, Russia\\
$^4$ ITEP, Bol. Cheremushkinskaya 25, Moscow, 117218, Russia
}

\date{Accepted XXX. Received YYY; in original form ZZZ}

\pubyear{2020}

\begin{document}
\label{firstpage}
\pagerange{\pageref{firstpage}--\pageref{lastpage}}
\maketitle

\begin{abstract}
We report the discovery of a nearby massive white dwarf with He-H atmosphere. The white dwarf is located at a distance of $74.5\pm0.9$\,pc. Its radius, mass, effective temperature, H/He ratio and age are $R=2500\pm100$\,km, $M=1.33\pm0.01$\,$\rm M_{\odot}$, $T_{\rm eff}=31200\pm 1200$\,K, ${\rm H/He}\sim0.1$ and $330\pm40$\,Myr, respectively. The observed spectrum is redshifted by $V_{\rm r}=+240\pm15$\,km\,s$^{-1},$ which is mostly attributed to the gravitational redshift. The white dwarf shows a regular stable photometric variability with amplitude $\Delta g\approx0.06^{\rm m}$ and period $P=353.456$~s suggesting rapid rotation. This massive, hot and rapidly rotating white dwarf is likely to originate from the merging of close binary white dwarf system that avoided explosion in a thermonuclear type Ia supernova at the Carboniferous Period of the Earth history. 
\end{abstract}



\section{Introduction}
The candidate source, Pan-STARRS 118732780118122536 (${\rm RA}=18^{\rm h}32^{\rm m}02.8^{\rm s},$ ${\rm Dec}=08\degr56^{\prime}36^{\prime\prime}$) -- hereinafter WD1832+089 -- came to our attention during our survey of dim blue nearby objects in Pan-STARRS DR1 catalogue \citep{PS-DR1}.
First we have selected all objects in a 100 pc sphere according to  Gaia DR2 catalogue \citep{GAIA-DR2}. After applying quality cuts and excluding objects in 20 degrees radius cone around the direction to the Galactic center, we have selected a subset of blue objects with colour index $(B-R)<0$ and cross-matched it with the Pan-STARRS DR1 catalogue using a one-arcsec cone search radius.
Using the blackbody approximation, we calculated the effective temperatures for cross-matched objects,  after that known parallaxes allowed us to estimate their radii.

The most extreme object, Pan-STARRS 118732780118122536, was found to have an effective temperature of $T_{\mr{eff}}\sim38 000$~K. At a distance of $d=74.5$~pc the corresponding  radius is  $R\sim(1800-1900)$~km. This source was also observed by the {\it GALEX} telescope \citep{GALEX}. Two {\it GALEX} measurements of the spectral flux densities in the far- and near-UV ranges perfectly match the blackbody spectrum with temperature estimated from the Pan-STARRS observations alone.

To elucidate the nature of the source, we performed additional spectroscopic and  photometric observations with the 2.5-m telescope of CMO SAI MSU (Caucasian Mountain Observatory of Sternberg Astronomical Institute of Moscow State University). The observations suggested that the object is a hot, massive, rapidly rotating white dwarf which likely originated from the merging of close binary white dwarfs
about 330 Myr ago that failed to explode as a type Ia supernova.

The paper is organized as follows.
In Section~\ref{sec:observations} we describe our observations, Sections~\ref{sec:results} contains the obtained results. In Section \ref{sec:discussion}  we discuss the properties of the source and 
summarize our findings.

\section{Observations}
\label{sec:observations}

The observations were carried out on the 2.5-m telescope of Caucasian Mountain Observatory of Sternberg Astronomical Institute  \citep{kornilov14}.
The spectroscopy with 
a spectral resolution of $R\sim 1500$ in the 3500-7400 {\AA} range was done using the recently commissioned double-beam TDS spectrograph\footnote{\url{http://lnfm1.sai.msu.ru/kgo/instruments/tds/}} with a slit width of one arcsec. The instrument has already proved its efficiency in spectroscopic observations of optically faint  \textit{Spektr-RG}/eROSITA X-ray sources (Dodin et al., 2020, Astron. Lett., in press).

Two spectra of the source were obtained: on March 27, 2020, with a total exposure of 7200 s, and on April 8, with a total exposure of 4800 s. 

The data reduction was performed in a standard way, including the dark subtraction, cleaning for the cosmic rays, 2D wavelength calibration with Ne-Kr-Pb arc lamp and the flat-field correction.
The radiation flux was integrated within a three-arcsec aperture, the background was estimated using the clean regions above and below the target and removed before the integration. The flux was calibrated using spectral standards from the ESO list\footnote{\url{https://www.eso.org/sci/observing/tools/standards.html}}. However, since we used a narrow slit, the absolute calibration was impossible due to the slit losses. The wavelength calibration was corrected using  the night-sky emission lines and was accurate within 0.2 {\AA} and 0.1 {\AA} in the blue and red beam of the TDS spectrograph, respectively. Finally, the wavelengths were transformed to the barycentric reference frame by using {\sc astropy} \citep{astropy} functions.

The preliminary photometric observations of the source carried out on March 27 (107 x 60 s images) and April 09 (275 x 30 s images), 2020 with SDSS $g'$ filter at the robotic 60-cm RC600 telescope of CMO SAI MSU revealed a fast variability. 
To confirm it,  the optical photometry on the 2.5-m telescope was performed during three nights: on April 19 (197 x 10 s), May 13 (413 x 10 s) and May 15 (327 x 10 s), 2020, using a mosaic CCD camera NBI with a standard SDSS $g'$ filter. 
Data  reduction  for NBI CCD included  bias  subtraction,  correction  for  non-linearity of each detector, flat-field correction, removal of background air-glow  emission  and  removing  cosmic-ray  hits. After the data reduction, a differential photometry of the WD was performed using the VaST package \citep{vast}.
To search for periodicity in series of observations taken at different dates, the moments of observations were converted to the Heliocentric Julian Dates.

\section{Results}
\label{sec:results}
\subsection{Effective temperature and radius of the white dwarf}

\begin{figure}
 \begin{center}
\includegraphics[scale=0.5]{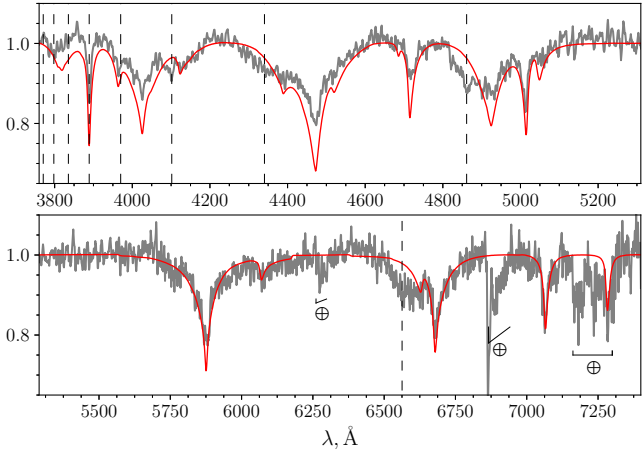}
 \end{center}
  \caption{The observed spectrum of the candidate (the grey line) and the synthetic spectrum of a DB white dwarf with $T_{\rm eff}=38\,000$\,K and $\log g=9.0$ (the red solid line). The position of hydrogen lines are marked with dashed lines. Both spectra are normalized to the continuum level. Telluric lines are marked with $\oplus.$}
 \label{fig:fig1}
\end{figure}

The spectrum of the WD candidate is shown in Fig. \ref{fig:fig1}. It corresponds to a helium white dwarf atmosphere with traces of hydrogen (DBA). For comparison, in this Figure with the red line we show the theoretical spectrum of a DB white dwarf with $T_{\rm eff}=38\,000$\,K and $\log g=9.0$ calculated by \citet{koester10}. The observed spectrum looks more shallow than the model one, demonstrates weak hydrogen lines and has a redshift of $\sim250$\,km\,s$^{-1}.$ 

\begin{figure}
 \begin{center}
\includegraphics[scale=0.5]{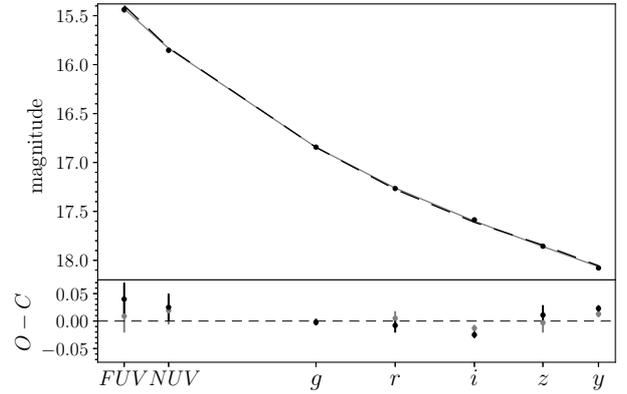}
 \end{center}
  \caption{Observed (dots) and theoretical (lines) magnitudes for various photometric bands (the upper panel) and their residuals (the lower panel). Grey colour is for Koestner's model ($\rm H/He=10^{-6}$), black colour is for {\sc tlusty} model with $\rm H/He=0.1.$ (see Table 1).}
 \label{fig:figSED}
\end{figure}

Having established the nature of the object as a DBA white dwarf, we can compare the obtained photometric observations with theoretical SEDs. 
To do this, we integrated Koester's SEDs with the corresponding transmission curves and fitted the theoretical magnitudes to the observed ones using the weighted least square method (see Fig. \,\ref{fig:figSED} and Table \ref{tab:tabSED}). Given the parallax value $13.43\pm0.17$ mas from Gaia DR2 catalogue, the best fit model parameters are $T_{\rm eff}=35\,000\pm600$\,K and $R=2580\pm50$\,km.
However, Koester's model spectra does not account for a presence of  hydrogen, which may lead to the effective temperature overestimation \citep{Eisenstein06}. To examine this effect, we calculated a set of model WD atmospheres and spectra in the LTE approximation using the {\sc tlusty} code \citep{Hubeny95,Hubeny03}. To check our calculations, we compared the model {\sc tlusty} spectra with Koester's spectra with the same parameters. The good agreement enabled us to use our models for the SED fitting. As expected, an increase in the hydrogen fraction in the WD atmosphere decreases the inferred $T_{\rm eff}$, however the WD radius remains constant ( within its determination errors ). The observed ratio between the depth of hydrogen and helium lines can be  obtained if the  hydrogen fraction is close to 10 per cent. Then the WD effective temperature and radius are $T_{\rm eff}=31200\pm1200$\,K  and $R=2520\pm100$\,km, respectively. The $\log g$ value only weakly affects the results, so in our model spectral calculations we have adopted $\log g=9.5$. 

\begin{table}
\caption{Observed and theoretical SED magnitudes for various photometric bands and models.} 
 \label{tab:tabSED} 
 \begin{center}
\begin{tabular}{ccccccc}
\hline
         & Observed     & Koester      & {\sc tlusty}       & {\sc tlusty}      \\
\hline         
$\log {\rm H}/{\rm He}$         &         & -6 & -6 & -1\\
$T_{\rm eff},$\,$10^{3}$K& & 35.0 & 36.2 & 31.2\\
$R,$\,km          &        & 2570 & 2530 & 2520\\
\hline
NUV    &$15.438\pm0.030$   &  15.429   &   15.415   & 15.398 \\
FUV    &$15.853\pm0.025$   &  15.834   &   15.835   & 15.828 \\
g      &$16.843\pm0.006$   &  16.845   &   16.839   & 16.845 \\
r      &$17.266\pm0.013$   &  17.261   &   17.274   & 17.274 \\
i      &$17.586\pm0.006$   &  17.599   &   17.617   & 17.611 \\
z      &$17.856\pm0.018$   &  17.859   &   17.843   & 17.845 \\
y      &$18.078\pm0.006$   &  18.065   &   18.054   & 18.055 \\
\hline 
\end{tabular}
\end{center}
\end{table}

\subsection{White dwarf mass and age} 

The high effective temperature and small radius of the WD as inferred from spectral fitting needs explanation. The small radius suggests a high WD mass $\sim 1.3~ \rm M_\odot$, about twice as heavy as the mean mass of observed DB WDs $\sim 0.67~ \rm M_\odot$ \citep{Blinnikov1994, Bergeron2011}.

\begin{figure}
 \begin{center}
\includegraphics[scale=0.5]{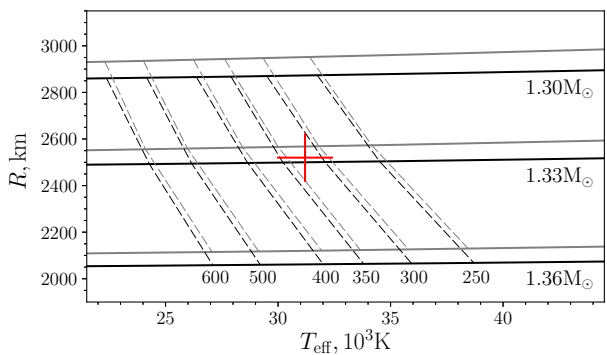}
 \end{center}
  \caption{Cooling curves for a WD with $M=1.30,$ 1.33, 1.36~$\rm M_{\odot}$ in the $T_{\rm eff}-R$ coordinates.
  The black lines are for pure He envelope, the grey lines are for He-H envelope. The dashed lines are isochrones for different ages, which are subscribed in Myr at the bottom of each line. The red cross represents parameters of WD1832+089 with their uncertainties.}
 \label{fig:figTR}
\end{figure}

Using our WD cooling code \citep{Blinnikov1993, Blinnikov1994, Popov2018}, we calculated cooling tracks for CO white dwarfs with masses 1.30, 1.33, 1.36~$\rm M_{\odot}$. Results slightly depend on the composition of WD envelope: the smallest radius is obtained in models with a pure helium atmosphere. Results for models with a helium layer $M_{\rm He}=2.4\times 10^{-2}$\,$\rm M_{\odot}$ are shown in Fig. \,\ref{fig:figTR} with black lines. For these models the observed effective temperature and radius corresponds to a white dwarf with mass  $M=1.33\pm0.01$\,$\rm M_{\odot}$ and age $330\pm40$\,Myr (see Fig. \,\ref{fig:figTR}). The adding of hydrogen to the helium envelope (with $M_{\rm H}=1.5\times10^{-4}$\,$\rm M_{\odot}$) slightly increases the mass estimate, but it still 1.33\,$\rm M_{\odot}$ within uncertainties (see Fig. \,\ref{fig:figTR}). We have also performed calculations for another bracketing case -- Si WD and found that the mass estimates does not depend on the WD core composition.
For such a massive white dwarf, a significant gravitational redshift of $230$\,km\,s$^{-1}$ is expected. It is very close to the observed value $\sim250$\,km\,s$^{-1}$ which corroborates the WD parameters derived from our observations.

\begin{figure}
 \begin{center}
\includegraphics[scale=0.5]{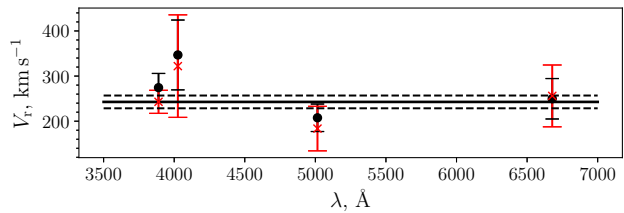}
 \end{center}
  \caption{Radial velocities of \ion{He}{i} lines measured in spectra obtained on March 27 (black dots) and April 8 (red crosses). The solid and dashed lines show the weighted average and corresponding uncertainty (the error of mean).}
 \label{fig:figVr}
\end{figure}

With found WD parameters, we can measure the spectral redshift more precisely by using the Koester's spectrum as a template.
To do this, we chose four helium lines and determined their spectral shift using the least square method by adjusting simultaneously the continuum level and the line depth.
We treated the two observed spectra separately, allowing for possible variations in radial velocities. The obtained results were found to be identical within the errors (see Fig. \,\ref{fig:figVr} and Table \,\ref{tab:radvel}). The weighted average of all measurements is $V_{\rm r}=+240\pm15$\,km\,s$^{-1}.$

\begin{table}
\caption{Radial velocities of \ion{He}{i} lines derived from the spectral fitting.} 
 \label{tab:radvel} 
 \begin{center}
\begin{tabular}{ccc}
\hline
Line & March 27, 2020        & April 8, 2020    \\
     &$V_{r},$\,km\,s$^{-1}$ &$V_{r},$\,km\,s$^{-1}$\\
\hline
\ion{He}{i}\,3889 & $274 \pm 31$ & $243 \pm 26$   \\
\ion{He}{i}\,4026 & $347 \pm 77$ & $322 \pm114$   \\
\ion{He}{i}\,5015 & $208 \pm 30$ & $184 \pm 49$   \\
\ion{He}{i}\,6678 & $250 \pm 45$ & $256 \pm 69$   \\
\hline
\end{tabular}
\end{center}
\end{table}

Part of this redshift can be attributed to the radial velocity of the star. We can not measure this velocity separately from the gravitational redshift, but we can estimate it from the proper motion, which is 9 mas~$\mathrm{yr}^{-1}$ and corresponds to 3\,km\,s$^{-1}$ at  distance 74.47\,pc. Therefore, it is unlikely that the radial velocity exceeds the estimated uncertainty of 15\,km\,s$^{-1}.$

\subsection{Photometric variability}

\begin{figure}
 \begin{center}
\includegraphics[scale=0.5]{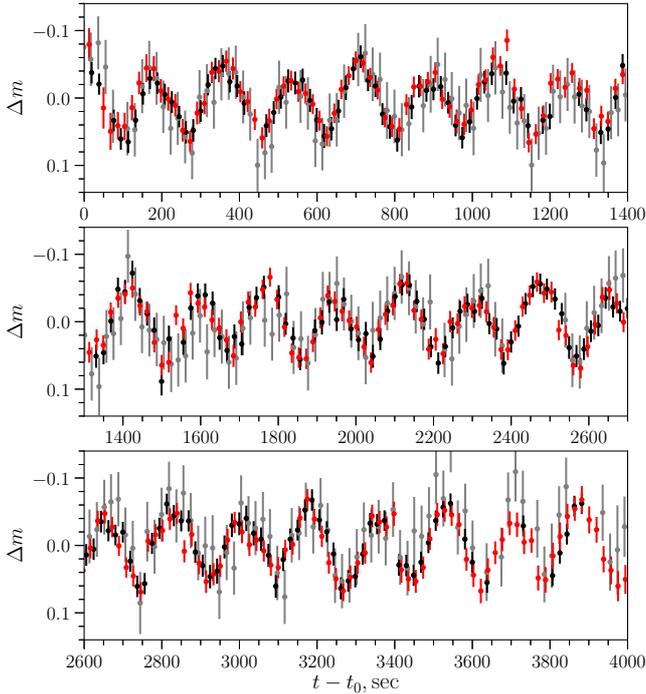}
 \end{center}
  \caption{Photometric data obtained on April 19, May 13 and May 15, 2020. For comparison, the light curves are shifted by an integer number of periods:
  the black dots are for $t_0={\rm JD}_0=2458959.475056$ (April 19),
  the grey dots are for $t_0={\rm JD}_0+5856P_0=2458983.431542$ (May 13),
  the red dots are for $t_0={\rm JD}_0+6356P_0=2458985.477007$ (May 15).
  The accepted period is $P_0=0.00409093^{\rm d}.$}
 \label{fig:figLC}
\end{figure}

Most of the photometric data obtained during three nights in April-May 2020 are presented in Fig.\,\ref{fig:figLC}. It can be seen that the shape of the light curve remains stable throughout the entire observation period (26 days) and demonstrates two maxima of different heights. The stability of the light curve shape and its phase suggests that this variability is caused by stellar rotation rather than WD pulsations.
\begin{figure}
 \begin{center}
\includegraphics[scale=0.5]{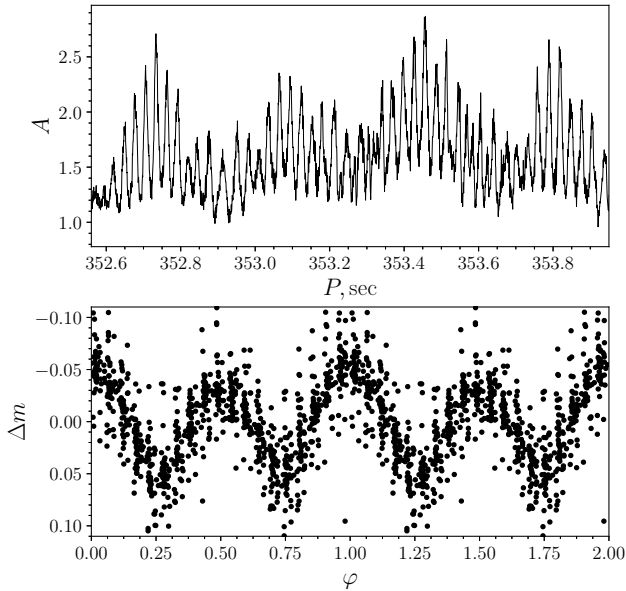}
 \end{center}
  \caption{The Lafler-Kinman periodogram (the upper panel) and phase curve (the bottom panel) corresponding to $P_0=353.456$ s and $\rm JD_0=2458959.475056$. }
 \label{fig:figLK}
\end{figure}

We searched for periodicity in the photometric light curves by using the 
Lafler-Kinman method \citep{LK65}. Several possible periods in the range 352--354 seconds were found (see Fig.\,\ref{fig:figLK}) with the strongest peak in the periodogram being at a period of 353.456 sec (0.00409093$^{\rm d}$).
This short period makes WD1832+089 the second fastest spinning isolated WD after SDSSJ125230.93-023417.72 with a rotation period of 317 seconds \citep{Reding2020}.

\section{Discussion  and conclusions}
\label{sec:discussion}

The analysis of our spectral observations suggests that the source is a hot massive DBA white dwarf
with $T_\mathrm{eff}=31200$~K and mass $M=1.33~ M_\odot$. 
There are two main scenarios of origin of ultramassive white dwarfs, like WD1832+089. First, such a WD could be an end product of the evolution of a single massive  ($\sim 9~\mathrm{M}_{\odot}$) star (assuming the solar chemical abundance). Such stars can complete their post-helium nuclear burning  as ONe WDs with masses above $\sim 1~\mathrm{M}_{\odot}$ \citep{Garcia-Berro1997,Woosley2015}. However, the cores of red giants are not expected to be rapidly rotating \citep{2012A&A...548A..10M}.

Another possible formation scenario of massive white dwarfs is a merger of two white dwarfs in a close binary system
\citep{Bergeron1991,Segretain1997}. This mechanism has gained a lot of attention due to its likely relevance to type Ia supernova explosions (the so-called double-degenerate scenario, see, e.g., \citet{2014LRR....17....3P}). Here, however, we are interested in a less cataclysmic outcome of the binary WD merging, when a single WD with mass not much lower then the total mass of the merging progenitor WDs is formed. Simulations of the WD merger process have been performed in the framework of smooth-particle hydrodynamics and are very complicated (e.g.\citep{Shen2012,Schwab2012,Zhu2013,Dan2014}). The end result of the merging mainly depends on  the total mass of the binary components, their mass difference (or the mass ratio), chemical compositions, and on whether the binary orbit had been synchronized before the merging or not. In certain setups, it is possible to avoid an explosive carbon ignition in the center which presumably leads to a type Ia supernova explosion. Instead, a slow carbon burning eventually transforms the merging object into a ONe WD \citep{Shen2012}, or no burning occurs at all and the final state of the merger remnant is a massive CO WD \citep{Dan2014}. In the WD merger scenario, the remnant is expected to be rapidly rotating because the orbital angular momentum of the merging binary is partially transferred to it.
The rapid $\sim 6$~min rotation of the massive and hot WD1832+089 found from our observations supports the merging scenario of its formation.

Scanning X-ray observations by the eROSITA telescope onboard of the \textit{Spektr-RG} satellite \citep{2012arXiv1209.3114M} gave only upper limits on the source flux in the soft X-ray energy range: $F_{0.3-2.0~\rm{keV}} \le 8\times10^{-14}~\mr{erg~cm^{-2}~s^{-1}}$ \citep{srg_lims}. At distance $d=74.5~\mr{pc}$, this corresponds to limits on the X-ray luminosity $L_X<5\times10^{28}~\mr{erg~s^{-1}}$. By assuming the possible accretion X-ray power, this allows us to constrain the accretion rate onto the WD $\dot{m}_\mathrm{acc}<7 \times 10^{10}~\mr{g~s^{-1}}$.
This is by an order of magnitude below the standard Bondi-Hoyle accretion rate from the ISM with density $\rho=10^{-24}$~{g~cm}$^{-3}$ for the fiducial WD velocity 10 km~s$^{-1}$. However, the object is located within the  Local Hot Bubble \citep{2009Ap&SS.323....1W} extending up to $\sim 100$~pc in the direction to the source \citep{2015ApJ...806..120S} and the Bondi-Hoyle accretion rate  strongly decreases in the rarefied ($n\sim 0.01$~cm$^{-3}$) hot ionized gas.
Clearly, deeper X-ray observations of the source would be valuable.

The lack of noticeable Zeeman splitting of \ion{He}{i} lines in the spectrum enables us to put upper limits on the strength of the WD surface magnetic field  $B\lesssim1~$MG. For this low field, no significant effects of the interaction of the possible WD magnetosphere with the surrounding plasma are expected. 

We conclude that our spectroscopic and photometric observations of the extreme blue object Pan-STARRS 118732780118122536 revealed its nature as a nearby hot ($T_\mathrm{eff}\approx 31200$~K), ultramassive ($M\sim 1.33~ M_\odot$), rapidly rotating ($P\approx 353$~s)  DBA white dwarf, which is most likely to be a remnant of the coalescence of a close binary white dwarf system that failed to explode as a type Ia supernova $\sim 330$~Myr ago at the Carboniferous Period.


\section*{Acknowledgements}

The work of MP is supported by the
Foundation for the Advancement of Theoretical Physics and Mathematics
``BASIS'' grant 18-1-2-51-1.
The work of AD, AB, SZh (observations and spectral modeling)   is supported by the Scientific School of the Moscow State University 'Physics of Stars, Relativistic Objects, and Galaxies'. AD acknowledges support from RSF grant 17-12-01241 regarding TDS data processing.
The authors acknowledge support from M.V.Lomonosov Moscow State University Program of Development in expanding the instrumentation base of the CMO SAI MSU.
The work of SB and KP (white dwarf cooling calculations and evolutionary considerations) is supported by the RSF grant 19-12-00229.  
This research has made use of NASA's Astrophysics Data System. 

\section*{Data availability}
The spectroscopic and photometric data used in this paper can be shared on request from the corresponding author.

\bibliographystyle{mnras}

\begin{thebibliography}{}
\makeatletter
\relax
\def\mn@urlcharsother{\let\do\@makeother \do\$\do\&\do\#\do\^\do\_\do\%\do\~}
\def\mn@doi{\begingroup\mn@urlcharsother \@ifnextchar [ {\mn@doi@}
  {\mn@doi@[]}}
\def\mn@doi@[#1]#2{\def\@tempa{#1}\ifx\@tempa\@empty \href
  {http://dx.doi.org/#2} {doi:#2}\else \href {http://dx.doi.org/#2} {#1}\fi
  \endgroup}
\def\mn@eprint#1#2{\mn@eprint@#1:#2::\@nil}
\def\mn@eprint@arXiv#1{\href {http://arxiv.org/abs/#1} {{\tt arXiv:#1}}}
\def\mn@eprint@dblp#1{\href {http://dblp.uni-trier.de/rec/bibtex/#1.xml}
  {dblp:#1}}
\def\mn@eprint@#1:#2:#3:#4\@nil{\def\@tempa {#1}\def\@tempb {#2}\def\@tempc
  {#3}\ifx \@tempc \@empty \let \@tempc \@tempb \let \@tempb \@tempa \fi \ifx
  \@tempb \@empty \def\@tempb {arXiv}\fi \@ifundefined
  {mn@eprint@\@tempb}{\@tempb:\@tempc}{\expandafter \expandafter \csname
  mn@eprint@\@tempb\endcsname \expandafter{\@tempc}}}

\bibitem[\protect\citeauthoryear{{Astropy Collaboration} et~al.,}{{Astropy
  Collaboration} et~al.}{2013}]{astropy}
{Astropy Collaboration} et~al., 2013, \mn@doi [\aap]
  {10.1051/0004-6361/201322068}, \href
  {https://ui.adsabs.harvard.edu/abs/2013A&A...558A..33A} {558, A33}

\bibitem[\protect\citeauthoryear{{Bergeron}, {Kidder}, {Holberg}, {Liebert},
  {Wesemael}  \& {Saffer}}{{Bergeron} et~al.}{1991}]{Bergeron1991}
{Bergeron} P.,  {Kidder} K.~M.,  {Holberg} J.~B.,  {Liebert} J.,  {Wesemael}
  F.,   {Saffer} R.~A.,  1991, \mn@doi [\apj] {10.1086/169972}, \href
  {https://ui.adsabs.harvard.edu/abs/1991ApJ...372..267B} {372, 267}

\bibitem[\protect\citeauthoryear{{Bergeron} et~al.,}{{Bergeron}
  et~al.}{2011}]{Bergeron2011}
{Bergeron} P.,  et~al., 2011, \mn@doi [\apj] {10.1088/0004-637X/737/1/28},
  \href {https://ui.adsabs.harvard.edu/abs/2011ApJ...737...28B} {737, 28}

\bibitem[\protect\citeauthoryear{{Blinnikov} \&
  {Dunina-Barkovskaya}}{{Blinnikov} \&
  {Dunina-Barkovskaya}}{1993}]{Blinnikov1993}
{Blinnikov} S.~I.,  {Dunina-Barkovskaya} N.~V.,  1993, Astronomy Reports, \href
  {https://ui.adsabs.harvard.edu/abs/1993ARep...37..187B} {37, 187}

\bibitem[\protect\citeauthoryear{{Blinnikov} \&
  {Dunina-Barkovskaya}}{{Blinnikov} \&
  {Dunina-Barkovskaya}}{1994}]{Blinnikov1994}
{Blinnikov} S.~I.,  {Dunina-Barkovskaya} N.~V.,  1994, \mn@doi [\mnras]
  {10.1093/mnras/266.2.289}, \href
  {https://ui.adsabs.harvard.edu/abs/1994MNRAS.266..289B} {266, 289}

\bibitem[\protect\citeauthoryear{{Chambers} et~al.,}{{Chambers}
  et~al.}{2016}]{PS-DR1}
{Chambers} K.~C.,  et~al., 2016, arXiv e-prints, \href
  {https://ui.adsabs.harvard.edu/abs/2016arXiv161205560C} {p. arXiv:1612.05560}

\bibitem[\protect\citeauthoryear{{Dan}, {Rosswog}, {Br{\"u}ggen}  \&
  {Podsiadlowski}}{{Dan} et~al.}{2014}]{Dan2014}
{Dan} M.,  {Rosswog} S.,  {Br{\"u}ggen} M.,   {Podsiadlowski} P.,  2014,
  \mn@doi [\mnras] {10.1093/mnras/stt1766}, \href
  {https://ui.adsabs.harvard.edu/abs/2014MNRAS.438...14D} {438, 14}

\bibitem[\protect\citeauthoryear{{Eisenstein} et~al.,}{{Eisenstein}
  et~al.}{2006}]{Eisenstein06}
{Eisenstein} D.~J.,  et~al., 2006, \mn@doi [\aj] {10.1086/504424}, \href
  {https://ui.adsabs.harvard.edu/abs/2006AJ....132..676E} {132, 676}

\bibitem[\protect\citeauthoryear{{Gaia Collaboration} et~al.,}{{Gaia
  Collaboration} et~al.}{2018}]{GAIA-DR2}
{Gaia Collaboration} et~al., 2018, \mn@doi [\aap]
  {10.1051/0004-6361/201833051}, \href
  {https://ui.adsabs.harvard.edu/abs/2018A&A...616A...1G} {616, A1}

\bibitem[\protect\citeauthoryear{{Garc{\'\i}a-Berro}, {Ritossa}  \&
  {Iben}}{{Garc{\'\i}a-Berro} et~al.}{1997}]{Garcia-Berro1997}
{Garc{\'\i}a-Berro} E.,  {Ritossa} C.,   {Iben} Icko J.,  1997, \mn@doi [\apj]
  {10.1086/304444}, \href
  {https://ui.adsabs.harvard.edu/abs/1997ApJ...485..765G} {485, 765}

\bibitem[\protect\citeauthoryear{{Gilfanov}, {Medvedev}  \&
  {Sunyaev}}{{Gilfanov} et~al.}{2020}]{srg_lims}
{Gilfanov} M.,  {Medvedev} P.,   {Sunyaev} R.,  2020, private communication

\bibitem[\protect\citeauthoryear{{Hubeny} \& {Lanz}}{{Hubeny} \&
  {Lanz}}{1995}]{Hubeny95}
{Hubeny} I.,  {Lanz} T.,  1995, \mn@doi [\apj] {10.1086/175226}, \href
  {https://ui.adsabs.harvard.edu/abs/1995ApJ...439..875H} {439, 875}

\bibitem[\protect\citeauthoryear{{Hubeny} \& {Lanz}}{{Hubeny} \&
  {Lanz}}{2003}]{Hubeny03}
{Hubeny} I.,  {Lanz} T.,  2003, in {Hubeny} I.,  {Mihalas} D.,   {Werner} K.,
  eds,  Astronomical Society of the Pacific Conference Series Vol. 288, Stellar
  Atmosphere Modeling. p.~51

\bibitem[\protect\citeauthoryear{{Koester}}{{Koester}}{2010}]{koester10}
{Koester} D.,  2010, \memsai, \href
  {https://ui.adsabs.harvard.edu/abs/2010MmSAI..81..921K} {81, 921}

\bibitem[\protect\citeauthoryear{{Kornilov} et~al.,}{{Kornilov}
  et~al.}{2014}]{kornilov14}
{Kornilov} V.,  et~al., 2014, \mn@doi [\pasp] {10.1086/676648}, \href
  {https://ui.adsabs.harvard.edu/abs/2014PASP..126..482K} {126, 482}

\bibitem[\protect\citeauthoryear{{Lafler} \& {Kinman}}{{Lafler} \&
  {Kinman}}{1965}]{LK65}
{Lafler} J.,  {Kinman} T.~D.,  1965, \mn@doi [\apjs] {10.1086/190116}, \href
  {https://ui.adsabs.harvard.edu/abs/1965ApJS...11..216L} {11, 216}

\bibitem[\protect\citeauthoryear{{Merloni} et~al.,}{{Merloni}
  et~al.}{2012}]{2012arXiv1209.3114M}
{Merloni} A.,  et~al., 2012, arXiv e-prints, \href
  {https://ui.adsabs.harvard.edu/abs/2012arXiv1209.3114M} {p. arXiv:1209.3114}

\bibitem[\protect\citeauthoryear{{Morrissey} et~al.,}{{Morrissey}
  et~al.}{2007}]{GALEX}
{Morrissey} P.,  et~al., 2007, \mn@doi [\apjs] {10.1086/520512}, \href
  {https://ui.adsabs.harvard.edu/abs/2007ApJS..173..682M} {173, 682}

\bibitem[\protect\citeauthoryear{{Mosser} et~al.,}{{Mosser}
  et~al.}{2012}]{2012A&A...548A..10M}
{Mosser} B.,  et~al., 2012, \mn@doi [\aap] {10.1051/0004-6361/201220106}, \href
  {https://ui.adsabs.harvard.edu/abs/2012A&A...548A..10M} {548, A10}

\bibitem[\protect\citeauthoryear{{Popov}, {Mereghetti}, {Blinnikov}, {Kuranov}
  \& {Yungelson}}{{Popov} et~al.}{2018}]{Popov2018}
{Popov} S.~B.,  {Mereghetti} S.,  {Blinnikov} S.~I.,  {Kuranov} A.~G.,
  {Yungelson} L.~R.,  2018, \mn@doi [\mnras] {10.1093/mnras/stx2910}, \href
  {https://ui.adsabs.harvard.edu/abs/2018MNRAS.474.2750P} {474, 2750}

\bibitem[\protect\citeauthoryear{{Postnov} \& {Yungelson}}{{Postnov} \&
  {Yungelson}}{2014}]{2014LRR....17....3P}
{Postnov} K.~A.,  {Yungelson} L.~R.,  2014, \mn@doi [Living Reviews in
  Relativity] {10.12942/lrr-2014-3}, \href
  {https://ui.adsabs.harvard.edu/abs/2014LRR....17....3P} {17, 3}

\bibitem[\protect\citeauthoryear{{Reding}, {Hermes}, {Vanderbosch}, {Dennihy},
  {Kaiser}, {Mace}, {Dunlap}  \& {Clemens}}{{Reding} et~al.}{2020}]{Reding2020}
{Reding} J.~S.,  {Hermes} J.~J.,  {Vanderbosch} Z.,  {Dennihy} E.,  {Kaiser}
  B.~C.,  {Mace} C.~B.,  {Dunlap} B.~H.,   {Clemens} J.~C.,  2020, \mn@doi
  [\apj] {10.3847/1538-4357/ab8239}, \href
  {https://ui.adsabs.harvard.edu/abs/2020ApJ...894...19R} {894, 19}

\bibitem[\protect\citeauthoryear{{Schwab}, {Shen}, {Quataert}, {Dan}  \&
  {Rosswog}}{{Schwab} et~al.}{2012}]{Schwab2012}
{Schwab} J.,  {Shen} K.~J.,  {Quataert} E.,  {Dan} M.,   {Rosswog} S.,  2012,
  \mn@doi [\mnras] {10.1111/j.1365-2966.2012.21993.x}, \href
  {https://ui.adsabs.harvard.edu/abs/2012MNRAS.427..190S} {427, 190}

\bibitem[\protect\citeauthoryear{{Segretain}, {Chabrier}  \&
  {Mochkovitch}}{{Segretain} et~al.}{1997}]{Segretain1997}
{Segretain} L.,  {Chabrier} G.,   {Mochkovitch} R.,  1997, \mn@doi [\apj]
  {10.1086/304015}, \href
  {https://ui.adsabs.harvard.edu/abs/1997ApJ...481..355S} {481, 355}

\bibitem[\protect\citeauthoryear{{Shen}, {Bildsten}, {Kasen}  \&
  {Quataert}}{{Shen} et~al.}{2012}]{Shen2012}
{Shen} K.~J.,  {Bildsten} L.,  {Kasen} D.,   {Quataert} E.,  2012, \mn@doi
  [\apj] {10.1088/0004-637X/748/1/35}, \href
  {https://ui.adsabs.harvard.edu/abs/2012ApJ...748...35S} {748, 35}

\bibitem[\protect\citeauthoryear{{Snowden}, {Koutroumpa}, {Kuntz}, {Lallement}
  \& {Puspitarini}}{{Snowden} et~al.}{2015}]{2015ApJ...806..120S}
{Snowden} S.~L.,  {Koutroumpa} D.,  {Kuntz} K.~D.,  {Lallement} R.,
  {Puspitarini} L.,  2015, \mn@doi [\apj] {10.1088/0004-637X/806/1/120}, \href
  {https://ui.adsabs.harvard.edu/abs/2015ApJ...806..120S} {806, 120}

\bibitem[\protect\citeauthoryear{{Sokolovsky} \& {Lebedev}}{{Sokolovsky} \&
  {Lebedev}}{2018}]{vast}
{Sokolovsky} K.~V.,  {Lebedev} A.~A.,  2018, \mn@doi [Astronomy and Computing]
  {10.1016/j.ascom.2017.12.001}, \href
  {https://ui.adsabs.harvard.edu/abs/2018A&C....22...28S/abstract} {22, 28}

\bibitem[\protect\citeauthoryear{{Welsh} \& {Shelton}}{{Welsh} \&
  {Shelton}}{2009}]{2009Ap&SS.323....1W}
{Welsh} B.~Y.,  {Shelton} R.~L.,  2009, \mn@doi [\apss]
  {10.1007/s10509-009-0053-3}, \href
  {https://ui.adsabs.harvard.edu/abs/2009Ap&SS.323....1W} {323, 1}

\bibitem[\protect\citeauthoryear{{Woosley} \& {Heger}}{{Woosley} \&
  {Heger}}{2015}]{Woosley2015}
{Woosley} S.~E.,  {Heger} A.,  2015, \mn@doi [\apj]
  {10.1088/0004-637X/810/1/34}, \href
  {https://ui.adsabs.harvard.edu/abs/2015ApJ...810...34W} {810, 34}

\bibitem[\protect\citeauthoryear{{Zhu}, {Chang}, {van Kerkwijk}  \&
  {Wadsley}}{{Zhu} et~al.}{2013}]{Zhu2013}
{Zhu} C.,  {Chang} P.,  {van Kerkwijk} M.~H.,   {Wadsley} J.,  2013, \mn@doi
  [\apj] {10.1088/0004-637X/767/2/164}, \href
  {https://ui.adsabs.harvard.edu/abs/2013ApJ...767..164Z} {767, 164}

\makeatother
\end{thebibliography}





\bsp	
\label{lastpage}
\end{document}